\documentclass[aps,prb,twocolumn,superscriptaddress,longbibliography]{revtex4-2}
\usepackage[colorlinks=true,linkcolor=blue,citecolor=blue]{hyperref}
\usepackage{amsfonts}
\usepackage{subfigure}
\usepackage{amsmath}
\usepackage{txfonts}
\usepackage{amssymb}
\usepackage{amsbsy}
\usepackage{epsfig}
\usepackage{graphicx}
\usepackage{epstopdf}
\usepackage{mathdots}
\usepackage{color}
\usepackage{cleveref}
\usepackage{bbm}
\usepackage{natbib}
\usepackage{url}
\usepackage[utf8]{inputenc}
\usepackage[makeroom]{cancel}
\usepackage{multirow}
\usepackage[table]{xcolor}
\usepackage{colortbl}

\newcommand{\be}{\begin{equation}}
\newcommand{\ee}{\end{equation}}

\newcommand{\bs}{\boldsymbol}

\newcommand{\sgn}{{\rm sgn}}

\begin{document}

\title{Cataloging topological phases of $\bs{N}$-stacked Su-Schrieffer-Heeger chains by a systematic breaking of symmetries}
\author{Aayushi Agrawal}
\email{p20170415@pilani.bits-pilani.ac.in}
\affiliation{Department of Physics, Birla Institute of Technology and Science, Pilani 333031, India}
\author{Jayendra N. Bandyopadhyay}
\email{jnbandyo@gmail.com}
\affiliation{Department of Physics, Birla Institute of Technology and Science, Pilani 333031, India}

\date{\today}

\begin{abstract}

Two-dimensional (2D) model of a weak topological insulator with $N$-stacked Su-Schrieffer-Heeger (SSH) chain is studied. This study starts with a basic model with all the fundamental symmetries (chiral, time-reversal, and particle-hole) preserved. Different topological phases are introduced in this model by systematically breaking the system's symmetries. The symmetries are broken by introducing different bonds (hopping terms) in the system. First, the chiral symmetry is broken by introducing hopping within each sub-lattice or intra-sub-lattice hopping, where the hopping strengths of the sub-lattices are equal in magnitudes but opposite in sign. Then, following Haldane, the time-reversal (TR) symmetry is broken by replacing the real intra-sub-lattice hopping strengths with imaginary numbers without changing the magnitudes. We find that breaking chiral and TR symmetries are essential for the weak topological insulator to be a Chern insulator. These models exhibit nontrivial topology with the Chern number $C = \pm 1$. The preservation of the particle-hole (PH) symmetry in the system facilitates an analytical calculation of $C$, which agrees with the numerically observed topological phase transition in the system. An interesting class of topologically nontrivial systems with $C=0$ is also observed, where the non-triviality is identified by quantized and fractional 2D Zak phase. Finally, the PH symmetry is broken in the system by introducing unequal amplitudes of intra-sub-lattice hopping strengths, while the equal intra-sub-lattice hopping strengths ensures the preservation of the inversion symmetry. We investigate the interplay of the PH and the inversion symmetries in the topological phase transition. A discussion on the possible experimental realizations of this model is also presented.

\end{abstract}

\maketitle

\section{\label{sec:Introduction}Introduction}

Topological insulators (TIs) \cite{Ryu2002,Moore2010,MZHasan2010,Topologicalinsulators2006,TopologicalTilen2017} are intriguing materials with wide technological applications in photonics, quantum computers, spintronics, topological electronics, etc \cite{Photonic2013,Topologicalinsulator2019,Topologicalelectronics2021}. Unlike ordinary insulators with an energy band gap in the bulk, a distinct feature of the TIs is the presence of edge states in the gap that connect the valence and conduction bands. This allows electronic charges to flow along the edges. The most basic prototype model for one-dimensional (1D) TIs is the Su-Schrieffer-Heeger (SSH) model \cite{Ryu_2010, asboth2016short}, which was proposed to model the electrical properties of a polyacetylene chain \cite{su1979solitons, heeger1988solitons}. The topological properties of a single SSH chain are measured by winding number \cite{Zak1989}. 

Various extensions of the 1D SSH model, such as ladder structures and two coupled-SSH chains, have been proposed \cite{Ladder-01,Ladder-02,Ladder-03,Ladder-04,Monkman2020,Borja_2022,Zhang2017,Sivan2022,Ladder-05}. An extension of the 1D SSH model (E-SSH) with the nearest-neighbor (NN) and the next-nearest-neighbor (NNN) hopping modulated by a cyclic parameter has been studied extensively \cite{li2014topological,2018EL_12437003L,Agrawal_2022,Xie2019}. The E-SSH model was considered as an effective 2D model, where the cyclic parameter plays the role of an additional synthetic dimension. For this effective 2D system, Chern number (C) characterizes the topological properties \cite{Chern1983} and it was observed that the presence of the NNN hopping leads to the topological phases with $C = \pm 1$. This study shows that the phase diagram of the E-SSH model could be mapped to the Haldane model \cite{li2014topological}. We have recently studied a Floquet version of this model, where we have observed Floquet-Bloch bands with very high Chern numbers \cite{Agrawal_2022}. 

A series of works have studied two versions of the 2D SSH model: First, a square lattice, where each unit cell contains four different atoms \cite{2DSSH2019,2DSSH_Li_2022}. Therefore, this model has {\it four} energy bands. An interesting feature of this 2D SSH model is that it shows nontrivial topology but with zero Berry curvature \cite{2DSSH_2017}. Second, a layered 2D SSH model is constructed by stacking $N$-number of SSH chains \cite{chen2018two,APalyi,WeylSSH-01,WeylSSH-02,WeylSSH-03}. Moreover, one can obtain a layered structure of the 2D SSH model by promoting the cyclic parameter of the E-SSH model to (quasi)momentum. In general, a $(d+1)$ dimensional layered structure can be constructed by stacking $d$ dimensional TIs. This layered material is categorized as a new class of TIs, called {\it weak} TIs. Since the stacking happens in some particular direction, one can expect edge modes only in the stacking direction. Therefore, unlike the four bands 2D SSH model, where four edge modes may appear at all four sides, the $N$-stacked SSH model has edge modes only at the two sides along the direction of stacking. This results in an anisotropy of the edge states in the system. 

Originally, this idea of stacking was incorporated in a generalized 3D quantum spin Hall (QSH) system \cite{KaneMele3D2007, Moore3D2007}, where they formed 3D weak TIs from the layered 2D QSH \cite{KaneGraphene2005, KaneMele2D2005}. They also showed two distinct classes: weak TIs and strong TIs, depending on the nature of their surface states \cite{WTI-01}. In this study, they observed that a small amount of disorder could destroy weak TIs and transform these into band insulators. A later study has demonstrated that the weak TIs are protected from any random disorder, provided this disorder does not break the TR symmetry and does not close the bulk energy gap \cite{WTI-02}. Thereafter, a new topological phase was discovered near the transition from weak TIs to strong TIs, called topological {\it semimetal} (TSM) \cite{TopSemi-01}. The TSMs are the phases where two (four) bands are separated by a finite band gap, however, there exist some points at the Fermi energy where both (four) bands have degeneracy \cite{TopSemi-02,TopSemi-03}. These degenerate band touching points are called Weyl (Dirac) nodes \cite{WeySemi-04, DiracSemi-05}.

Recently, a couple of extensive studies have found that, by varying the system parameters, the $N$-stacked SSH model makes a transition from trivial insulator to weak topological insulator via topological semimetal state \cite{chen2018two,APalyi}. The experimental realization of this model has also been proposed \cite{Yang2022}. Since this model is chiral symmetric, we calculate the winding number as a topological invariant by dimension reduction. Until now, the 2D $N$-stacked SSH model has been studied to preserve all the fundamental symmetries. Therefore, the above studies raise a natural question about the fate of the $N$-stacked SSH chains with various broken symmetries. The primary goal of this study is to investigate the impact of the broken symmetries on the $N$-stacked SSH model and catalog their interesting topological phases.

In this paper, we have considered a basic model of $N$-stacked half-filled SSH chains. This basic model preserves chiral, time-reversal (TR), and particle-hole (PH) symmetries like in the single SSH chain. The presence of these symmetries in the system suggests that its Hamiltonian satisfies the following conditions in quasi-momentum space or $\bs{k}$-space:
\begin{equation}
\begin{array}{ccc}
\mathcal{P}^{-1} \mathcal{H}(k_x,k_y)\, \mathcal{P} & = & -\mathcal{H}(-k_x,-k_y),\\
\mathcal{T}^{-1} \mathcal{H}(k_x,k_y)\, \mathcal{T} & = & \mathcal{H}(-k_x,-k_y),\\
\mathcal{C}^{-1} \mathcal{H}(k_x,k_y)\, \mathcal{C} & = & -\mathcal{H}(k_x,k_y),
\end{array}
\label{Eq-Symmetries}
\end{equation}
where $\mathcal{P}$, $\mathcal{T}$, and $\mathcal{C}$ respectively represent chiral, TR, and PH operations. Based on these three fundamental symmetries, a periodic table of the topological materials was proposed to classify them \cite{Symmetry-01, Symmetry-02, Symmetry-03}. A major part of this paper, preserving the PH symmetry in the system, investigate the topological properties of the system in the presence or absence of the chiral and the TR symmetries. These symmetries are systematically preserved or broken by allowing or restricting various hopping in the system. The TR symmetry is broken by considering hopping strengths with imaginary amplitude. Later, we also study the role of  the PH symmetry on the topology of the system.
 
In this model of the $N$-stacked SSH chains, we have placed individual SSH chains along the $x$-direction, whereas in the $y$-direction, we have stacked $N$ number of identical SSH chains. Here we have proposed different versions of the $N$-stacked SSH chains depending on the various hopping terms in the system. We consider two specific cases for each of these versions: in one case, the individual SSH chains is topologically trivial (the winding number $w = 0$), and in the other case, the individual chain is topologically nontrivial ($w \neq 0$). We classify these different versions into two classes of Hamiltonians: 
\begin{enumerate} 
\item System with chiral symmetry. Because of this symmetry, off-diagonal terms appear in the Hamiltonian represented in the quasi-momentum space. 
\item System without chiral symmetry. Along with the off-diagonal terms, now diagonal or mass terms appear in the Hamiltonian.  
\end{enumerate}
For the latter class, we have considered two sub-classes based on the presence and absence of the TR symmetry.  
 
This paper is organized in the following manner. Section \ref{sec:Case-I} presented the model 2D $N$-stacked SSH chains with chiral symmetry. In the next section, Sec. \ref{sec:Case-II} discusses the model Hamiltonian with broken chiral symmetry. The next section, Sec. \ref{sec:Case-II}, is divided into two subsections based on the presence and absence of the TR-symmetry in the system. Here, we also extensively discuss nontrivial topological properties of all these models. Our numerical results show that these models show two types of nontrivial topology: with Chern number $C \neq 0$ and with $C=0$. In Sec. \ref{sec:Results}, an analytical calculation of the Chern number and the topological phase transition are presented. This calculation agrees well with the numerics. In Sec. \ref{sec:ZeroCN}, we focus on the topologically nontrivial models with $C = 0$. This nontrivial property is identified by calculating the quantized and fractional Zak phases. We also show the results with broken PH symmetry in Sec. \ref{sec:PH-broken} for both the previous cases discussed in Secs. \ref{sec:Case-II}A and \ref{sec:Case-II}B. In Sec. \ref{sec:Exp-aspects}, we also propose an experiment to simulate 2D $N$-stacked SSH chains. Finally, we summarize in Sec. \ref{sec:Final Remarks} and provide a future outlook in Sec. \ref{sec:Outlook}.

\section{\label{sec:Case-I} $\bs{N}$-stacked SSH chains with chiral symmetry}

\begin{figure}[b]
\centering
\includegraphics[width=6cm,height=3.5cm]{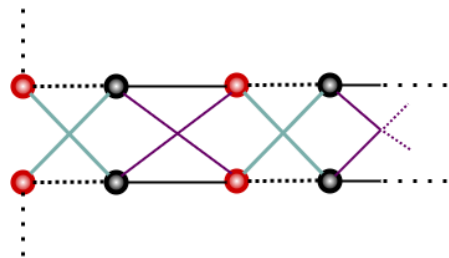}
\caption{\label{Fig-Primary_Model} Schematic representation of the 2D $N$-stacked SSH chains with chiral symmetry is presented. The red and black colored spheres, respectively, represent sites of $A$ and $B$ sub-lattices. The thick and dotted black bonds represent the inter-cell and intra-cell hopping within a chain, respectively. The neighboring chains are connected by light blue and purple dotted bonds.}
\end{figure}

In this section, we study the $N$-stacked SSH chains model with chiral symmetry. This is the basic model whose structure is shown in Fig. (\ref{Fig-Primary_Model}). Here, $N$ number of identical SSH chains are stacked along the $y$-direction, forming a 2D square/rectangular lattice model. Each SSH chain is a standard one with two sites per unit cell or one dimer per unit cell. Sites $A$ and $B$ are respectively represented by the red and black circles in the figure. Here we see that the {\it spin-less} electrons in this system can hop only to adjacent or nearest neighbor sites through the bonds represented by the dotted and solid black lines along the $x$-direction. The electrons are allowed to hop only between neighboring chains, which are shown by the light blue and purple lines. This 2D lattice has only two sub-lattices. Therefore, like an SSH chain, this system has two energy bands.

The real-space Hamiltonian of this model is given as
\begin{equation}
\begin{split}
H_{N-SSH} &= \sum_{n_x,n_y} \left[(1-\eta) \, c^{\dagger A}_{n_x,n_y} c^{B}_{n_x,n_y} + (1+\eta) \, c^{\dagger A}_{n_x+1,n_y} c^{B}_{n_x,n_y}\right]\\
& - \frac{\delta}{2} \sum_{n_x,n_y}\left[c^{\dagger A}_{n_x+1,n_y} c^{B}_{n_x,n_y+1} + c^{\dagger A}_{n_x+1,n_y+1} c^{B}_{n_x,n_y} \right]\\
&+ \frac{\delta}{2} \sum_{n_x,n_y} \left[c^{\dagger A}_{n_x,n_y} c^{B}_{n_x,n_y+1} + c^{\dagger A}_{n_x,n_y+1} c^{B}_{n_x,n_y} \right] + {\rm h. c.},
\end{split}
\label{Eq-H_SSH_real}
\end{equation}
where $c_{n_x,n_y}^A$ (and $c_{n_x,n_y}^B$) and $c^{\dagger A}_{n_x,n_y}$ (and $c^{\dagger B}_{n_x,n_y}$) are the fermionic annihilation and creation operators corresponding to the sub-lattice $A$ (and $B$). The first and the second terms of the Hamiltonian represent intra-cell and inter-cell hopping. The parameter $\eta$ fixes the relative strengths of these hopping, and thus it decides the topological property of the individual SSH chain. 
The last two terms of the Hamiltonian describe the coupling between two adjacent SSH chains, and the parameter $\delta$ decides the strength of these hopping. Note that all the inter-chain hopping strengths are considered equal here. For the case of the full periodic boundary condition (PBC), i.e., the PBC is considered along both $x$- and $y$-directions, we have to stack $N$ number of SSH chains on a toroidal surface. Correspondingly, the quasi-momentum space will also be a 2D toroidal surface, defined in $\bs{k} = (k_x,\,k_y)$ space.  For the PBC, the above Hamiltonian in $\bs{k}$-space is expressed by substituting the real-space fermionic operators $\{c_{n_x, n_y}\}$ with its $\bs{k}$-space representation as
\begin{equation}
c_{n_x,n_y} = \frac{1}{N_x N_y} \sum_{k_x,k_y} \, e^{i (n_x k_x + n_y k_y)} \, \tilde{c}_{k_x,k_y}, 
\end{equation}
where $N_x$ is the total number of unit cells (or dimers) along $x$-direction and $N_y$ is the number of SSH chains stacked along $y$-direction. We thus get the $\bs{k}$-space Hamiltonian as
\begin{equation}
H_{N-SSH} = \sum_{\bs{k}} \Psi_{\bs{k}}^{\dagger} \, \mathcal{H}_{N-SSH}(\bs{k}) \, \Psi_{\bs{k}},
\end{equation}
where $\Psi_{\bs{k}} = \left[c_{\bs{k}}^{A} ~~ c_{\bs{k}}^{B}\right]^T$ are the Nambu spinors and the Hamiltonian kernel or the Bloch Hamiltonian $\mathcal{H}_{N-SSH}(\bs{k})$ is expressed as 
\begin{equation}
\begin{split}
\mathcal{H}_{N-SSH}(\bs{k}) &= \bs{h}(\bs{k}) \bs{\cdot} \bs{\sigma},~~{\rm where} \\ h_x(\bs{k}) &= \big[(1 + \cos k_x) + (1-\cos k_x) (\delta \cos k_y - \eta) \big]\\
h_y(\bs{k}) &= \big[(1+\eta) - \delta \cos k_y \big] \, \sin k_x,
\end{split}
\label{Eq-H_SSH_k}
\end{equation}
were $\sigma_\alpha$'s with $\alpha = \{x,y,z\}$ are Pauli's pseudo-spin matrices. Two energy bands of this system are obtained from the eigenvalues of the above Bloch Hamiltonian and these are
\begin{equation}
\begin{split}
E_{\pm}(\bs{k}) &= \pm \sqrt{h_{x}(\bs{k})^2 \, + \, h_{y}(\bs{k})^2 }\\
&= \pm \, \sqrt{2} \big[ \, (1+\cos k_x) + (1-\cos k_x) (\delta \cos k_y - \eta)^2 \, \big]^{1/2}.
\label{Eq-E_k form}
\end{split}
\end{equation}
Here, we consider two cases of the $N$-stacked SSH chains model: in one case, we set $\eta > 0$ to make the individual SSH chain topologically nontrivial; and in the other case, when $\eta < 0$, the individual SSH chain is topologically trivial. The presence of the edge states in the bulk gap characterizes the topological property of the system. The edge states are observed for the open boundary conditions (OBC) only. For better visibility of the edge states, here we consider energy bands under {\it partial} open boundary conditions (POBCs), when at a time the OBC is considered only along one direction, and the PBC is considered along the other direction. Geometrically, this suggests that the $N$-stacked SSH chains are now placed on a cylinder, where the cylinder's axis is along $x$-direction in one case and along $y$-direction in the other case. Under these two different POBC cases, the Hamiltonians become
\begin{equation}
\begin{split}
&\mathcal{H}_{N-SSH}(k_x) = \\ & \left[\Bigl\{\left(1-\eta\right) \, + \, (1+\eta) \cos k_x \Bigr\} \sigma_x + (1+\eta) \sin k_x \sigma_y\right] \otimes \mathbbm{1}_{Ny}\\ 
& + \frac{\delta}{2} \left[(1-\cos k_x) \sigma_x - \sin k_x \sigma_y\right] \otimes \sum_{n_y} (c_{n_y}^\dagger c_{n_y+1} + {\rm h.c.}) 
\end{split}
\label{H_kx}
\end{equation}
and
\begin{equation}
\begin{split}
&\mathcal{H}_{N-SSH}(k_y) = \sum_{n_x} \left[(1-\eta) \, c_{n_x}^{\dagger A} c_{n_x}^B + (1+\eta) \, c_{n_x+1}^{\dagger A} c_{n_x}^B\right] \\
& + \delta \cos k_y \sum_{n_x} \left[c_{n_x}^{\dagger A} c_{n_x}^B - c_{n_x+1}^{\dagger A} c_{n_x}^B\right] + {\rm h.c.}\\
& = \sum_{n_x} \left[\left(1-\eta + \delta \cos k_y\right) \, c_{n_x}^{\dagger A} c_{n_x}^B + \left(1+\eta - \delta \cos k_y \right) \, c_{n_x+1}^{\dagger A} c_{n_x}^B\right] \\
& + {\rm h.c.},
\end{split}
\label{H_ky}
\end{equation}
where $\mathcal{H}_{N-SSH}(k_x)$ is the Hamiltonian corresponding to the POBC case with the PBC is considered only along the $x$-direction; and $\mathcal{H}_{N-SSH}(k_y)$ represents the POBC Hamiltonian with the PBC is considered only along the $y$-direction. In other words, we can say that $\mathcal{H}_{N-SSH}(k_x)$ describes a model of $N$-stacked SSH chains placed on the surface of a cylinder, whose axis is along the $y$-direction; whereas $\mathcal{H}_{N-SSH}(k_y)$ describes the same 2D model on the surface of a cylinder, whose axis is along the $x$-direction. Here, $\mathbbm{1}_{Ny}$ is a $N \times N$ identity matrix. 
\begin{figure}[t]
\includegraphics[width=8cm,height=5cm]{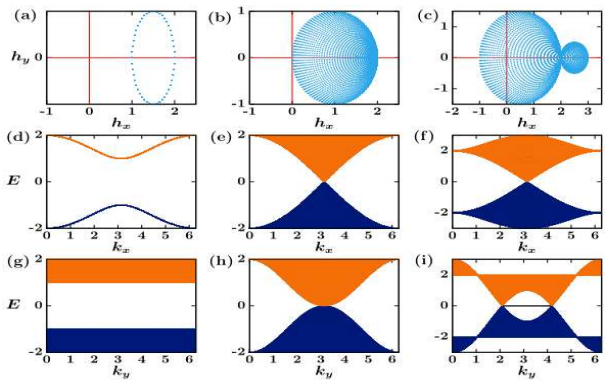}
\caption{In (a)-(c), the closed curves in the $h_x-h_y$ plane are presented for the model proposed in Fig. \ref{Fig-Primary_Model} for $N_y=200$ topologically trivial stacked SSH chains with $100$ dimers in every single chain (i.e., $N_x=200$). Here, we consider three different cases: $\delta = 0.0$, $\delta = 0.5$ and $\delta =1$ with $\eta = - 0.5$. In (d)-(f), corresponding energy bands are plotted as a function of $k_x$ for the POBC case, where PBC is considered only along $x$-direction and OBC is considered along $y$-direction. Similarly, in (g)-(i), energy bands are plotted as a function of $k_y$, where PBC is considered only along the $y$-direction and OBC is considered along the $x$-direction.} 
\label{Fig-Primary_ModelTri}
\end{figure}
\begin{figure}[t]
\includegraphics[width=8cm,height=5cm]{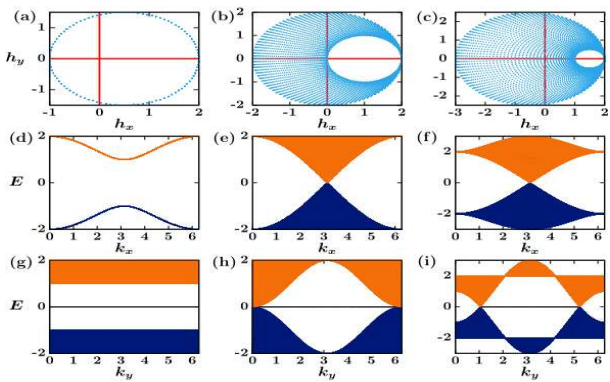}
\caption{The same results are presented here as in Fig. (\ref{Fig-Primary_ModelTri}), but here all the individual SSH chain is topologically nontrivial. Here, we again consider the same three different values of the parameter $\delta$ ($= 0.0, 0.5\, \mbox{and}\, 1.0$) and fix the other parameter $\eta = 0.5$.}
\label{Fig-Primary_ModelNonTri}
\end{figure}

In Eq. (\ref{H_ky}), the Hamiltonian $\mathcal{H}_{N-SSH}(k_y)$ mimics a single SSH chain, where $k_y$ plays the role of a parameter that modulates the hopping amplitudes. For each value of $k_y$, this Hamiltonian represents an SSH chain. Consequently, the defined weak topological invariant is independent of $k_y$, and that can be calculated only along the $x$-direction. If we compare with a single SSH chain, this system is topologically nontrivial when $\left(1+ \eta -\delta \cos k_y \right) > \left(1- \eta +\delta \cos k_y \right)$. This imposes the following condition on $k_y$ in the first Brillouin zone (BZ): 
\begin{equation}
\cos^{-1} \left(\frac{\eta}{\delta}\right) < k_y < 2\pi- \cos^{-1} \left(\frac{\eta}{\delta}\right).
\label{Eq-k_y-cond}
\end{equation}
In Figs. \ref{Fig-Primary_ModelTri}(a)-(c) and \ref{Fig-Primary_ModelNonTri}(a)-(c), the Hamiltonian kernel $\bs{h}(\bs{k})$ is shown in the $h_x-h_y$ plane for individual trivial and nontrivial SSH chain, respectively for three different values of $\delta$. Fig. \ref{Fig-Primary_ModelTri} (a) and \ref{Fig-Primary_ModelNonTri} (a) are plotted for $\delta=0$, i.e., there is no interchain hopping. Therefore, it behaves like a single SSH chain. The closed curve encloses the origin only when the individual SSH chain is nontrivial. The next two figures are plotted for non-zero values of $\delta$. Here, in contrast to the single SSH chain, it has a finite number of circles, which either enclose the origin or pass through the origin or not encloses the origin depending on the above condition of $k_y$. This is clearly observed when the energy bands are plotted as a function of $k_y$ under POBCs. 

We show the energy bands corresponding to the Hamiltonian $\mathcal{H}_{N-SSH}(k_x)$ and $\mathcal{H}_{N-SSH}(k_y)$ in Fig. \ref{Fig-Primary_ModelTri}(d)-(f) and Fig. \ref{Fig-Primary_ModelTri}(g)-(i), respectively. Here, we consider the individual SSH chain to be topologically trivial. We observe that, as the interchain hopping $\delta$ increases, the 2D $N$-stacked SSH model constructed with the topologically trivial individual SSH chains starts exhibiting nontrivial nature. 

The energy band diagrams for the case when the individual SSH chain is nontrivial are shown in Fig. \ref{Fig-Primary_ModelNonTri}(d)-(i). For this case, the 2D $N$-stacked SSH model shows the zero energy edge states for all values of $\delta$. However, the length of the zero energy states is decided by the above condition on $k_y$. The region where the zero energy edge states are presented is known as ``Edge Brillouin zone (EBZ)". The edge states in the EBZ are bridged by two bulk nodes and this explains the topological semimetal behavior of the system \cite{APalyi}. For $\delta = 1$, besides the shifting of the band-touching points, we do not see any qualitative difference in the bands in Figs. \ref{Fig-Primary_ModelTri}(i) and \ref{Fig-Primary_ModelNonTri}(i). For this system, the phase transitions occur in the following parametric way:
\begin{equation}
\begin{array}{ccl}
 \eta < -\delta &:~& {\rm Trivial}\,\, {\rm insulator},\\
-\delta < \eta < \delta &:~& {\rm Topological}\,\, {\rm semimetal},\\
\eta > \delta &:~& {\rm Weak}\,\, {\rm topological}\,\, {\rm insulator}.
\end{array}
\end{equation}
In this study, we choose the parameters such that the system has edge states irrespective of the nature of the individual SSH chain, whether it is trivial or nontrivial.

\section{\label{sec:Case-II} Generalized $\bs{N}$-stacked SSH chains with broken chiral symmetry}

The chiral symmetry is broken in the $N$-stacked SSH chains by introducing bonds within the sub-lattices, leading to the intra-sub-lattice hopping. The Chern number is a topological invariant for systems with broken chiral symmetry. Moreover, here we have considered two sub-classes of systems depending on the presence and absence of the TR symmetry. Following Haldane's approach \cite{Haldane}, the TR symmetry is broken by introducing intra-sub-lattice hopping with imaginary strength. This section is divided into two subsections. In subsections \ref{With-TR} and \ref{Without-TR}, we have respectively discussed the topological properties of the system with and without the TR symmetry, while the chiral symmetry is broken for both cases. Note once again that the PH symmetry is preserved.

\subsection{\label{With-TR} Topological phases with TR symmetry}

As a target of designing $N$-stacked SSH models of nontrivial topology with the Chern number $C \neq 0$, we introduce additional intra-sub-lattice hopping in our basic model. This hopping introduces a mass term in the $\bs{k}$-space Hamiltonian. Consequently, this term breaks the chiral symmetry and lifts the degeneracy of the two bands, which results in an opening of the band gap. The broken chiral symmetry is the cause of the existence of non-zero and non-degenerate energy edge states. Here, we set $\gamma_A = -\gamma_B = \gamma$ to preserve the PH symmetry in the system, where $\gamma_A\,(\gamma_B)$ is $A\,{\rm to}\,A\,(B\,{\rm to}\,B)$ sub-lattice hopping strength. Note that here the intra-sub-lattice hopping strengths are equal in magnitude, but opposite in sign. As a consequence, the energy spectrum is symmetric about $E=0$. For the condition $\gamma_A = \gamma_B$, the spectrum does not remain symmetric about $E=0$ due to the broken PH symmetry, but the inversion symmetry (discussed later in Sec. \ref{sec:PH-broken}) is preserved in the system. This model preserves the TR symmetry due to the real value of the chiral symmetry-breaking terms in the Hamiltonian.

\begin{figure}[t]
\centering
\includegraphics[width=8cm,height=8cm]{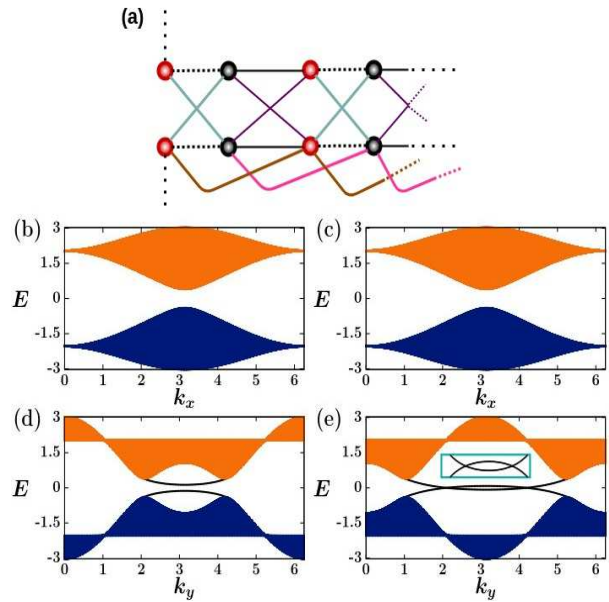}
\caption{A schematic diagram of a 2D $N$-stacked SSH model with broken chiral symmetry is shown in subfigure (a). Here, the intra-sub-lattice hopping strengths are real numbers and are shown by the solid-brown and magenta bonds. The corresponding energy bands under the POBCs are shown in subfigure (b)-(e). In subfigures (b) and (c), energy bands are shown as a function of $k_x$, where the PBC is considered along the $x$-direction and the OBC is considered along the $y$-direction. In subfigure (b), the individual SSH chain is trivial ($\eta = -0.5$); and in subfigure (c), each SSH chain is nontrivial ($\eta = 0.5$). For these two values of the parameter $\eta$, energy bands are shown as a function of $k_y$ in subfigures (d) and (e), respectively. In these cases, the PBC is considered along the $y$-direction, and the OBC is considered along the $x$-direction. Here, we set the parameters $\delta = 1.0$ and $\gamma = 0.2$.}
\label{Fig-E-Model_PT}
\end{figure}

We introduce NNN intra-chain (i.e., along the $x$-direction) bonds with real hopping strengths while keeping the original bonds intact. These bonds give additional hopping within an SSH chain from its $A$ site (or $B$ site) of a unit cell to the $A$ site (or $B$ site) of the adjacent unit cell. The schematic representation of this model is presented in Fig. \ref{Fig-E-Model_PT}(a). For this model, the Hamiltonian in the $\bs{k}$-space is given as:
\begin{equation}
\mathcal{H}_{\bs{k}} = \mathcal{H}_{N-SSH}(\bs{k}) +  2 \gamma \cos k_x \, \sigma_z.
\label{Eq-Re-x-hop}
\end{equation}
Here, the first term at the right side $\mathcal{H}_{N-SSH}(\bs{k})$ is given in Eq. (\ref{Eq-H_SSH_k}). The second term appears due to the newly introduced NNN bonds, where we set the intra-sub-lattice long-range hopping strength $\gamma = 0.2$. In Figs. \ref{Fig-E-Model_PT}(b)-(e), 
the energy bands are presented for the POBC. Here the system satisfies the PBC along the $y$-direction, but the OBC along the $x$-direction. The corresponding POBC Hamiltonian is 
\be
\mathcal{H}(k_y) = \mathcal{H}_{N-SSH}(k_y) + \gamma \sum_{n_x} \left[\left( c_{n_x}^{\dagger A} c_{n_x+1}^A - c_{n_x}^{\dagger B} c_{n_x+1}^B\right) + h.c.\right]
\ee
Here, $\mathcal{H}_{N-SSH}(k_y)$ is already given in Eq. \eqref{H_ky}. Figs. \ref{Fig-E-Model_PT}(b) and  \ref{Fig-E-Model_PT}(d) show the energy bands with the POBC for the Hamiltonian $\mathcal{H}(k_y)$, when the individual chain of the stacked SSH chains is topologically trivial (winding number $w=0$). On the other hand, Figs. \ref{Fig-E-Model_PT}(c) and (e) show the energy bands, when the individual SSH chain is topologically nontrivial ($w \neq 0$). 

In Fig. \ref{Fig-E-Model_PT}(d), we see the presence of the edge states at the band gap. These states are not connecting the valence and the conduction band. Therefore, the Chern number of this system is $C=0$. The energy band diagram of this system is similar to the bands shown in Ref. \cite{li2014topological}, where an SSH chain with an additional synthetic dimension was studied. For the other case, as shown in Fig. \ref{Fig-E-Model_PT}(e), an edge state emanating from one band forms a single lobe by crossing the edge state emanating from the other band twice and finally enters into the same band from where it was emanated. The edge state emanating from the other band shows the same property.  Here again, the Chern number $C=0$. The $C = 0$ is also observed for both cases by analytical means in Sec. \ref{sec:Results}. Even though for these cases $C=0$, the presence of the edge states indicates nontrivial topology. This nontriviality will be revealed by the calculation of the 2D Zak phase in Sec. \ref{sec:ZeroCN}.

\subsection{\label{Without-TR} Topological phases without TR symmetry}

In the previous subsection \ref{With-TR}, we observed that the TR-symmetric $N$-stacked SSH chains without chiral symmetry could not have a nontrivial topology with a non-zero Chern number. Here, we study the $N$-stacked SSH chains with broken chiral symmetry and TR symmetry. The chiral symmetry was broken earlier by introducing intra-sub-lattice hopping with real hopping strengths. Here, we simultaneously break the chiral and the TR symmetries by replacing all the intra-sub-lattice real hopping strengths with imaginary amplitudes. In experiments, the imaginary hopping amplitude can be realized by applying a magnetic field-like gauge field in the direction perpendicular to the plane of the 2D lattice. However, following Haldane \cite{Haldane}, we consider intra-sub- lattice imaginary hopping amplitude, which gives the gauge-field-like effect even without any external physical field. We again set the intra-sub-lattice hopping strength $\gamma_A = -\gamma_B = i \gamma$ to preserve the PH symmetry. Due to the broken TR symmetry, this system can now be transformed into a Chern insulator, a TR symmetry broken topological insulator with a non-zero Chern number.
	
\subsubsection{Model 1}

\begin{figure}[t]
\includegraphics[width=8cm,height=8cm]{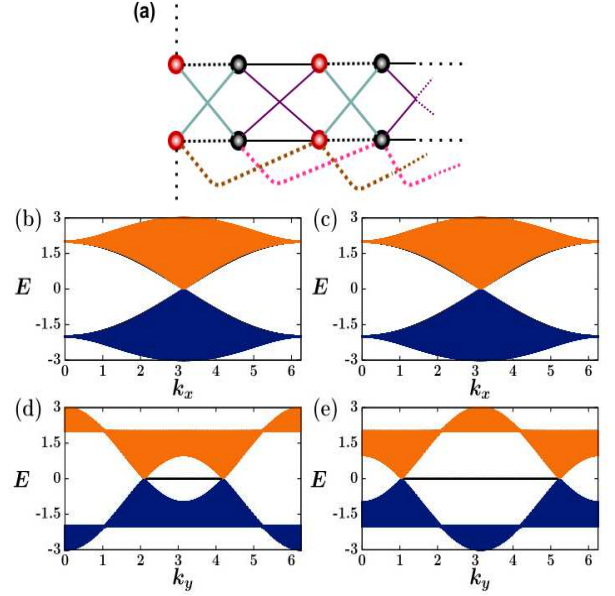}
\caption{Model 1: The subfigure (a) shows a model which is identical to the model presented in Fig. \ref{Fig-E-Model_PT}(a). However, here the strengths of all the intra-sub-lattice hopping are imaginary (shown by the dotted bonds), breaking the TR-symmetry in the system. The other subfigures show the energy bands of the system under the POBCs. In subfigures (b) and (c), energy bands with the PBC only along $x$-direction are presented. Here, in (b), the individual SSH chain is topologically trivial with $\eta = -0.5$; whereas, in (c), the individual chain is topologically nontrivial with $\eta = 0.5$. Similarly, in subfigures (d) and (e), the energy bands are presented respectively for the same values of the parameter $\eta$, but here individual SSH chain with PBC is considered only along $y$-direction. Here, we set $\delta = 1$ and $\gamma = 0.2$.}
\label{Fig-Im-x-hop}
\end{figure}

First, we study the system with NNN intra-chain hopping of imaginary amplitude from $A$ (or $B$) sites to $A$ (or $B$) sites of the nearest unit cell. The schematic diagram of this model is shown in Fig. \ref{Fig-Im-x-hop}(a). The Hamiltonian of this system in the real space (lattice space) is of the form
\begin{equation}
\begin{split}
&H_{GN-SSH} = H_{N-SSH} \\ &- i \gamma \, \sum_{n_x, n_y} \, \big(\, c^{\dagger A}_{n_x,n_y} c^{A}_{n_x+1,n_y} - c^{\dagger B}_{n_x,n_y} c^{B}_{n_x+1,n_y} \,- {\rm h.c.} \, \big). 
\end{split}
\end{equation}
The expression of the Hamiltonian $H_{N-SSH}$ is given in Eq. \eqref{Eq-H_SSH_real}. The above Hamiltonian in $\bs{k}$-space becomes
\begin{equation}
\mathcal{H}_{\bs{k}} = H_{N-SSH}(\bs{k}) + 2 \gamma \sin k_x \, \sigma_z,
\label{Eq-Im-x-hop}
\end{equation}
where the Hamiltonian $H_{N-SSH}(\bs{k})$ is given in Eq. \eqref{Eq-H_SSH_k}. Here again we set $\gamma = 0.2$. For this case, we present the band diagrams in Fig. \ref{Fig-Im-x-hop}(b)-(e) under POBCs. The corresponding Hamiltonians are
\begin{equation}
\begin{split}
&\mathcal{H}(k_x) = \mathcal{H}_{N-SSH}(k_x) + (2 \gamma \sin k_x) \, \sigma_z \otimes \mathbbm{1}_{Ny}\\
{\rm and}~\\
&\mathcal{H}(k_y) = \mathcal{H}_{N-SSH}(k_y) - i \gamma \sum_{n_x} \left(c_{n_x}^{\dagger A} c_{n_x+1}^A - c_{n_x}^{\dagger B} c_{n_x+1}^B - {\rm h. c.}\right). 
\end{split}
\label{Eq-Im-hop01}
\end{equation}
Figures \ref{Fig-Im-x-hop}(b) and \ref{Fig-Im-x-hop}(d) show energy bands for the case when the individual SSH chain is topologically trivial. On the other hand, Figs. \ref{Fig-Im-x-hop}(c) and \ref{Fig-Im-x-hop}(e) show energy bands for the case when the individual SSH chain is topologically nontrivial. Here, when the PBC is considered only along $x$-direction, we observe that the system is gapless and the bands are touching at $k_x = \pi$. However, when the PBC is considered only along the $y$-direction, the two bands touch at $k_y = \pi/3$ and $k_y = 2\pi/3$. These suggest that the system has a pair of  Dirac points at $(\pi,\,\pi/3)$ and $(\pi,\,2\pi/3)$. At both the Dirac points, the term appears in this model due to the broken chiral and TR symmetries vanishes. Therefore, like the basic model, here we also do not see any opening of band gap. Moreover, this system has a pair of degenerate zero energy edge states. We have calculated the Chern number of the lower band of this system and found $C=0$. Therefore, we now proceed to the next model, where {\it inter-chain} NNN hopping with imaginary amplitude is considered.

\subsubsection{Model 2}

\begin{figure}[b]
\includegraphics[width=8cm,height=8cm]{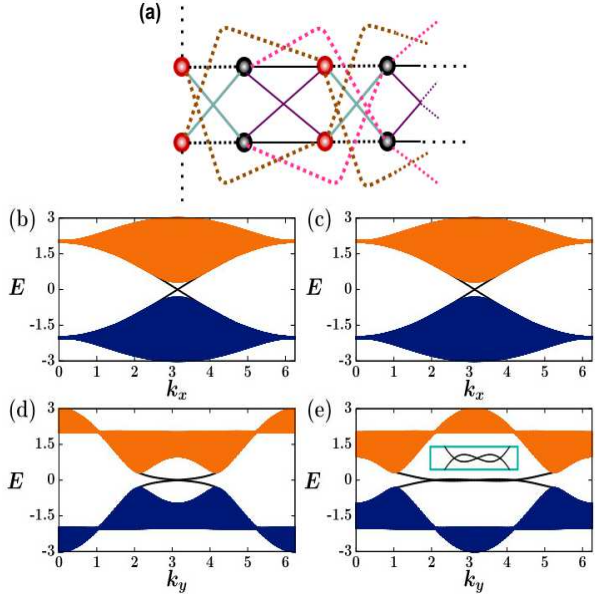}
\caption{Model 2: In subfigures (a), two different types of inter-chain hopping terms are considered, which are shown by brown (A to A) and magenta (B to B) bonds. These intra-sub-lattice hopping strengths are imaginary numbers (represented by the dotted bonds). The imaginary hopping amplitude breaks the TR-symmetry in the system. The remaining subfigures (b)-(e) show the energy bands of this model under POBCs same as for the cases described in Figs. \ref{Fig-Im-x-hop}(b)-(e).}
\label{Fig-Im-diag-hop}
\end{figure}

The schematic representation of this model is shown in Fig. \ref{Fig-Im-diag-hop}(a). Here, $A$ (or $B$) sites of a SSH chain are connected to the $A$ (or $B$) sites of the nearest unit cell of the neighboring SSH chain. In real space, the corresponding Hamiltonian under tight-binding condition is given as
\begin{equation}
\begin{split}
&H_{GN-SSH} = H_{N-SSH}\\ &- i \gamma \, \sum_{n_x, n_y} \, \big(\, c^{\dagger A}_{n_x,n_y} c^{A}_{n_x+1,n_y+1} - c^{\dagger B}_{n_x,n_y} c^{B}_{n_x+1,n_y+1} \,- {\rm h.c.} \, \big). 
\end{split}
\end{equation}
The above Hamiltonian under PBC can be represented in the $\bs{k}$-space as
\begin{equation}
\mathcal{H}_{\bs{k}} = H_{N-SSH}(\bs{k}) + 2 \gamma \, \sin (k_x + k_y) \, \sigma_z,
\label{Eq-Im-y-hop}
\end{equation}
where, as earlier, we set $\gamma = 0.2$. For this model, we show the band diagrams in Figs. \ref{Fig-Im-diag-hop}(b)-(e) under the POBCs and the corresponding Hamiltonians are
\begin{equation}
\begin{split}
\mathcal{H}(k_x) &= \mathcal{H}_{N-SSH}(k_x) + \left(\gamma \sin k_x\right) \, \sigma_z \otimes \sum_{n_y}
\left(c_{n_y}^\dagger c_{n_y+1} + {\rm h.c.} \right)\\ & - i \left(\gamma \cos k_x \right)\,\sigma_z \otimes \sum_{n_y} \left(c_{n_y}^\dagger c_{n_y+1} - {\rm h.c.}\right) \\
{\rm and}~~~~&\\ 
\mathcal{H}(k_y) &= \mathcal{H}_{N-SSH}(k_y) - i \gamma \sum_{n_x}\left[ \left(c_{n_x}^{\dagger A} c_{n_x+1}^A - c_{n_x}^{\dagger B} c_{n_x+1}^B\right) e^{-i k_y} - {\rm h. c.}\right]. 
\end{split}
\label{Eq-Im-hop02}
\end{equation}
Here, the energy bands for the case when the individual SSH chain is topologically trivial are shown in Figs. \ref{Fig-Im-diag-hop}(b) and (d). On the other hand, Figs. \ref{Fig-Im-diag-hop}(c) and (e) show the same for the case when the individual chain is topologically nontrivial. In Figs. \ref{Fig-Im-diag-hop}(b) and (c), the edge states with single crossing is observed when the energy bands are presented for the Hamiltonian $\mathcal{H}(k_x)$. However, in Figs. \ref{Fig-Im-diag-hop}(d) and (e), we respectively observe edge states with single and triple crossings (shown in the inset figure) in the energy bands of $\mathcal{H}(k_y)$. These band diagrams indicate that the system may exhibit nontrivial topological properties. We verify this by calculating the Chern number of the system and obtain $C = 1$ for both the cases. This study reveals that the system shows nontrivial topology, when the edge states cross odd number of times. 

In case of the previous model, the edge states were only observed for the POBC case with the PBC along the $y$-direction and the OBC along the $x$-direction. In this model, the edge states are observed along both the directions. This suggests that this system behaves like a true 2D model with two-bands, without any effect of the stacking in a particular direction. This system is a Chern insulator with $C=1$. These results motivate us to study the next model with $A$ to $A$ (or $B$ to $B$) NN inter-chain hopping of imaginary amplitude along the vertical or the $y$-direction.  

\subsubsection{Model 3}

\begin{figure}[t]
\includegraphics[width=8cm,height=8cm]{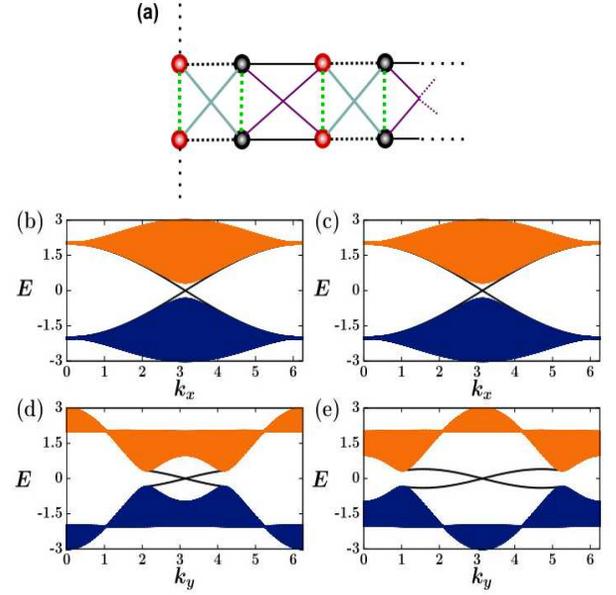}
\caption{Model 3: The subfigure (a) shows that here we consider a different version of $N$-stacked SSH chains with broken chiral and TR-symmetry. This model has intra-sub-lattice hopping along the vertical direction shown by the dotted green bonds. Here, once again, the remaining subfigures show the energy bands of this model under the POBCs for the cases identical to Figs. \ref{Fig-Im-x-hop}(b)-(e).}
\label{Fig-Im-y-hop}
\end{figure}

The schematic diagram of this system is shown in Fig. \ref{Fig-Im-y-hop}(a). The real-space Hamiltonian of this system is given as
\begin{equation}
\begin{split}
&H_{GN-SSH} = H_{N-SSH}\\ &- i \gamma \, \sum_{n_x, n_y} \, \big(\, c^{\dagger A}_{n_x,n_y} c^{A}_{n_x,n_y+1} - c^{\dagger B}_{n_x,n_y} c^{B}_{n_x,n_y+1} \,- h.c. \, \big), 
\end{split}
\end{equation}
and the corresponding Hamiltonian under the PBC is given in $\bs{k}$-space as
\begin{equation}
\mathcal{H}_{\bs{k}} = H_{N-SSH}(\bs{k}) + 2 \gamma \sin k_y \, \sigma_z,
\label{Eq-Im-diag-hop}
\end{equation}
where we set $\gamma = 0.2$. The band diagrams are shown in Figs. \ref{Fig-Im-y-hop}(b)-(e) for the POBC Hamiltonians
\begin{equation}
\begin{split}
&\mathcal{H}(k_x) = \mathcal{H}_{N-SSH}(k_x) - \left(i \gamma\right) \sigma_z \otimes \sum_{n_y} \left(c_{n_y}^\dagger c_{n_y+1} - {\rm h.c.} \right)\\ 
&{\rm and}\\
&\mathcal{H}(k_y) = \mathcal{H}_{N-SSH}(k_y) + 2 \gamma \sin k_y \sum_{n_x} \left(c_{n_x}^{\dagger A} c_{n_x}^A - c_{n_x}^{\dagger B} c_{n_x}^B\right). 
\end{split}
\label{Eq-Im-y-hop03}
\end{equation}
Similar to the previous cases, Figs. \ref{Fig-Im-y-hop}(b) and \ref{Fig-Im-y-hop}(d) show the energy bands, when all the individual SSH chain is topologically trivial. In Figs. \ref{Fig-Im-y-hop}(c) and \ref{Fig-Im-y-hop}(e), we have shown energy bands, where the individual SSH chain is topologically nontrivial. In all these figures, we observe edge states with single crossing only. Earlier, we have commented that Model 2 exhibits topologically nontrivial properties due to the odd number of crossings. In this model, we see single crossing of the edge states in the energy bands, which is also odd number of crossing. Therefore, we expect non-zero Chern number for this model. Our numerical calculation indeed finds $C = -1$. Like Model 2, again this model behaves like a true 2D Chern insulator with edge states along both the directions. Since, Model 2 and Model 3 show nontrivial topological properties with the Chern number of opposite signs, a natural question is whether the combination of these two models form a topologically trivial system. Thus we now study a combination of these two models.

\subsubsection{Model 4}

\begin{figure}[b]
\includegraphics[width=8cm,height=8cm]{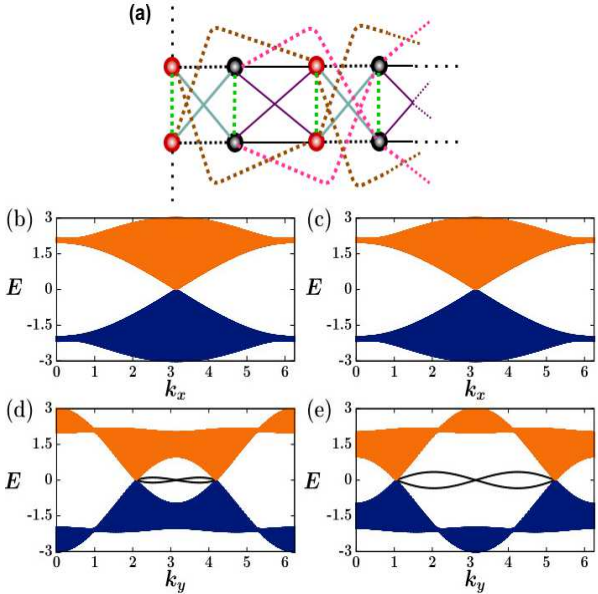}
\caption{Model 4: The schematic diagram presented in subfigure (a) shows that this model combines Model 2 and Model 3, as presented in Figs. (\ref{Fig-Im-diag-hop}) and (\ref{Fig-Im-y-hop}). The remaining subfigures show the energy bands of this model under POBCs, and their description is once again identical to Figs. \ref{Fig-Im-x-hop}(b)-(e).}
\label{Fig-Im-diag-y-hop}
\end{figure}

Finally, we consider a system having NNN diagonal hopping and NN vertical hopping among neighboring chains as depicted in Fig. \ref{Fig-Im-diag-y-hop}(a). Moreover, here we assume the strength of both the hopping as imaginary numbers. The corresponding real-space Hamiltonian is given as
\begin{equation}
\begin{split}
&H_{GN-CSSH}(n_x,n_y) = H_{N-CSSH}(n_x,n_y) - i \gamma \, \sum_{n_x, n_y} \, \big(\, c^{\dagger A}_{n_x,n_y} c^{A}_{n_x,n_y+1} \\
&+ c^{\dagger A}_{n_x,n_y} c^{A}_{n_x+1,n_y+1} -c^{\dagger B}_{n_x,n_y} c^{B}_{n_x,n_y+1} - c^{\dagger B}_{n_x,n_y} c^{B}_{n_x+1,n_y+1} \,- {\rm h.c.} \, \big).
\end{split}
\end{equation}
For this case, the $\bs{k}$-space Hamiltonian is 
\begin{equation}
\mathcal{H}_{\bs{k}} = H_{N-SSH}(\bs{k}) + 2 \gamma \, [\sin k_y + \sin (k_x + k_y) ] \, \sigma_z,
\label{Eq-Im-diag-y-hop}
\end{equation}
where we set $\gamma = 0.2$. The corresponding POBC Hamiltonians are
\begin{equation}
\begin{split}
&\mathcal{H}(k_x) = \mathcal{H}_{N-SSH}(k_x) + \gamma \sin k_x\, \sigma_z \otimes \sum_{n_y} \left(c_{n_y}^\dagger c_{n_y+1} + {\rm h.c.} \right)\\ &- i \gamma (1 + \cos k_x) \sigma_z \otimes \sum_{n_y} \left(c_{n_y}^\dagger c_{n_y+1} - {\rm h.c.} \right)\\
&{\rm and}\\
&\mathcal{H}(k_y) = \mathcal{H}_{N-SSH}(k_y) + 2 \gamma \sin k_y \sum_{n_x} \left(c_{n_x}^{\dagger A} c_{n_x}^A - c_{n_x}^{\dagger B} c_{n_x}^B\right)\\
&- i \gamma \sum_{n_x}\left[ \left(c_{n_x}^{\dagger A} c_{n_x+1}^A - c_{n_x}^{\dagger B} c_{n_x+1}^B\right) e^{-i k_y} - {\rm h. c.}\right].
\end{split} 
\label{Eq-Im-y-hop03}
\end{equation}
The energy bands corresponding to these Hamiltonians are presented in Figs. \ref{Fig-Im-diag-y-hop}(b)-(e). These figures show that the energy bands are gapless. Similar to Figs. \ref{Fig-E-Model_PT}(d) and \ref{Fig-E-Model_PT}(e), the Chern number of this model is expected to be $C=0$ due to the even number of crossings, and we have indeed found that by numerical calculation. Later, we shall also find the same by analytical calculation in Sec. \ref{sec:Results}. However, the presence of the edge states in this model is again indicating nontrivial topology, which is revealed in Sec. \ref{sec:ZeroCN} by the calculation of the 2D Zak phase.

Here we consider all the inter-chain hopping strengths are equal. However, for the unequal strengths of the interchain NNN {\it diagonal} hopping (dotted red and brown lines in the figure) and the NN hopping along the vertical direction (dotted green lines in the figure), this model is topologically nontrivial with $C = \pm 1$. The sign of the Chern number is dependent on the relative strength $\Gamma \equiv \gamma_1-\gamma_2$ of these two inter-chain hopping, where $\gamma_1$ is the inter-chain NNN diagonal hopping strength, and $\gamma_2$ is the strength of the inter-chain NN hopping along the vertical direction. When the NNN diagonal hopping strength is stronger than the NN vertical one, i.e., $\Gamma > 0$, the Chern number is $C = 1$. For the opposite case, when $\Gamma < 0$, we find $C = -1$. For both these cases, i.e., for $|\Gamma| >0$, we expectedly observe a gap in the bulk part of the energy bands of Model 4. We have already shown the gapless spectrum of Model 4 in Fig. \ref{Fig-Im-diag-y-hop} for $\Gamma = 0$, i.e. when $\gamma_1 = \gamma_2$. This implies $\Gamma = 0$ is the transition point for the topological phases with $C = 1$ to $C = -1$.

\subsection{A summary of the results presented in this section}

In all the cases discussed in this section, we observe that when the system has either NN vertical or NNN diagonal inter-chain hopping with imaginary amplitude, the systems exhibit nontrivial topology with the Chern number $C = \pm 1$. However, if we consider both these hopping in the system, we need unequal hopping strengths to get nontrivial topology with $C = \pm 1$. Furthermore, we observe that a system becomes topologically nontrivial with $C = \pm 1$ when the edge states emanating from the valence and conduction bands cross each other an odd number of times and connect the two bands. Compared with the energy band properties of the TR-symmetric systems with broken chiral symmetry, we observe according to the expectation that when the edge states cross each other an even number of times, the edge states do not connect the valence and the conduction bands. For these cases, we numerically find $C=0$. However, in these models, the presence of the edge states indicates their nontrivial topology. In the next section, Sec. \ref{sec:Results}, we have calculated the Chern number of these models analytically, and these agree with the numerics. Moreover, in Sec. \ref{sec:ZeroCN}, the nontrivial topology of the systems with the Chern number $C=0$ cases will be studied by the calculation of the 2D Zak phase as a topological invariant.

\section{\label{sec:Results} Chern number and phase diagram: An analytical calculation}

We have extensively studied different cases of the $N$-stacked SSH chains. In this study, we numerically find two cases showing nontrivial topology with Chern number $C = \pm 1$. The Chern number of the other cases was $C=0$. However, these models have shown their nontrivial nature by exhibiting edge states. In the previous sections, we calculated the Chern number numerically by integrating the Berry curvature over the first Brillouin zone \cite{Note}. In this section, we focus on the analytical calculation of the Chern number. Moreover, here we have shown phase diagrams of the topological transition.

\subsection{Chern number calculation}

Instead of calculating the Chern number via Berry curvature, we follow an alternate formula to calculate the Chern number \cite{Chernanalytic2012,Chernanalytic2022}. Here, instead of integrating the Berry curvature over the Brillouin zone, one needs to calculate summation of a quantity at all the Dirac points $D_i$, and the formula is given as 
\begin{equation}
C = \frac{1}{2} \sum_{\bs{k} \epsilon D_i} \, \sgn\left[\partial_{k_x} \bs{h(k)} \, \times \, \partial_{k_y} \bs{h(k)}\right]_z \, \sgn \left[h_z(\bs{k})\right].
\label{Eq-Chern_analytical}
\end{equation}
Here, we assume that the Hamiltonian in the $\bs{k}$-space is of the form $\mathcal{H}_{\bs{k}} = \bs{h(k) \cdot \sigma}$. If we substitute $\bs{h(k)}$ of the Hamiltonians considered in this paper in the above equation, we get the expression for the Chern number as
\begin{equation}
C = \frac{1}{2} \sum_{\bs{k} \epsilon D_i} \big[ \sgn\{-\delta \sin k_y (1+\eta - \delta \cos k_y) (1-\cos k_x)\} \, \sgn (h_z)\big].
\label{Eq-Chern_analytical-02}
\end{equation}

\begin{widetext}
\begin{table*}[t]
\centering
\begin{tabular}{|c|c|c|c|c|c|c|}
\hline
Symmetries & Section & Model \# & $h_z(\bs{k})$ & $h_z(\bs{k})\big|_{\bs{k} = D_1}$ & $h_z(\bs{k})\big|_{\bs{k} = D_2}$ & C\\
\hline\hline
$\mathcal{P}, \mathcal{C}, \mathcal{T}$ & Sec. \ref{sec:Case-I}& $-$ & $0$ & $0$ & $0$ & $0$\\
\hline $\mathcal{P}, \bcancel{\mathcal{C}}, \mathcal{T}$ & Sec. \ref{With-TR} & 1 & $2\gamma \cos k_x$ & $-2\gamma$ & $-2\gamma$ & $0$\\ \cline{3-7}
\hline
\multirow{4}{*}{$\mathcal{P}, \bcancel{\mathcal{C}}, \bcancel{\mathcal{T}}$} & \multirow{4}{*}{Sec. \ref{Without-TR}}& 1 & $2\gamma \sin k_x$ & $0$ & $0$ & $0$\\ \cline{3-7}
& &  $\bs{2}$ &  $2\gamma \sin (k_x + k_y)$ &  $ \mp 2\gamma \sqrt{1 - (\eta/\delta)^2}$ &  $\pm2\gamma \sqrt{1 - (\eta/\delta)^2}$ & $\bs{+1^{\bs{\mbox{*}}}}$\\ \cline{3-7}
 & & $\bs{3}$ & $2\gamma \sin k_y$ & $ \pm 2\gamma \sqrt{1 - (\eta/\delta)^2}$ & $\mp 2\gamma \sqrt{1 - (\eta/\delta)^2}$ & $\bs{-1^{\bs{\mbox{*}}}}$\\ \cline{3-7}
& & 4(a) & \multirow{2}{*}{} $\gamma_1 = \gamma_2$, $2\, [\gamma_1 \sin (k_x + k_y) + \gamma_2 \sin k_y]$ & $0$ & $0$ & $0$\\ \cline{3-7}
& & {\bf 4(b)} & $\gamma_1 \neq \gamma_2$, $2\, [\gamma_1 \sin (k_x + k_y) + \gamma_2 \sin k_y]$ & $ \pm 2(\gamma_2-\gamma_1) \sqrt{1 - (\eta/\delta)^2}$ & $\mp 2(\gamma_2-\gamma_1) \sqrt{1 - (\eta/\delta)^2}$ &         $\bs{-1^{\bs{\mbox{*}}}}$\\ 
\hline
\end{tabular}
\caption{A summary of the Chern number calculation using the analytical expression given in Eq. \eqref{eq:formula} is presented for all the systems studied in this paper. The extreme left column of the table shows the systems' presence and absence of different symmetries. Here, $\bcancel{\mathcal{C}}$ and $\bcancel{\mathcal{T}}$ denote respectively the broken chiral and the broken TR symmetries in the system. However, for all the systems, the PH symmetry $\mathcal{P}$ is preserved. $^{\bs{\mbox{*}}}$The sign of the Chern number can also be opposite if some system parameters change the sign. Topological phases of the systems with broken PH symmetry are not summarized here.}
\label{table}
\end{table*} 
\end{widetext}

\begin{figure}[b]
\includegraphics[width=8.5cm,height=5cm]{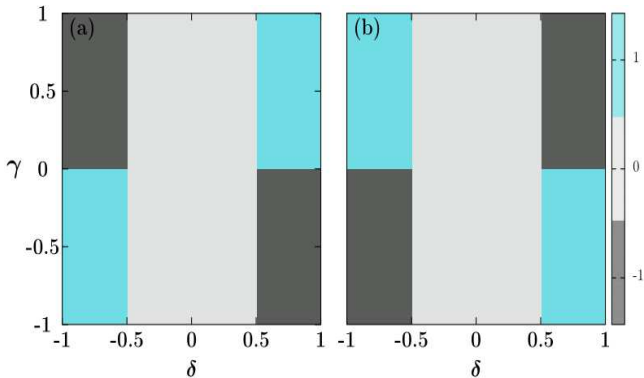}
\caption{The subfigures (a) and (b) show the topological phase transitions in the models presented in Figs. \ref{Fig-Im-diag-hop}(a) and \ref{Fig-Im-y-hop}(a), respectively. Here, the Chern number is calculated as a function of the parameters $\gamma$ and $\delta$ for a fixed value of $\eta = 0.5$. Varying the parameters $\gamma$ and $\delta$, three different topological phases with $C = 0, \pm1$ are observed.}
\label{Fig-Phase-diag} 
\end{figure}

At the Dirac points, any system with PH symmetry has doubly degenerate zero-energy states. These Dirac points are calculated by setting first $h_x(\bs{k}) = h_y(\bs{k}) = 0$, and then nullify $h_z(\bs{k})$ by tuning the system parameters. Following this, we get the Dirac points for all the models at the same place in the first Brillouin zone, and these are at
\be 
D_1 : \left[\pi, \cos^{-1} \left(\frac{\eta}{\delta}\right)\right]~~ {\rm and}~~ D_2 : \left[\pi, 2 \pi - \cos^{-1} \left(\frac{\eta}{\delta}\right)\right].
\label{eq:Dirac_points}
\ee
The above relation shows that the Dirac points can exist (i.e., bands can touch each other) only when $|\delta| > |\eta|$. We observe that these Dirac points are not on the high-symmetric path \cite{Feng_2021}. Therefore, by changing a system parameter, the Dirac points can be moved anywhere in the BZ. Substituting the Dirac points in Eq. \eqref{Eq-Chern_analytical-02} with the condition $\left|\frac{\eta}{\delta}\right| < 1$ and assuming $\delta > 0$ without losing any generality, we obtain a simplified expression of the Chern number for our systems as
\be
\begin{split}
C &= -\frac{1}{2} \sgn\left(\pm \sqrt{1 - \frac{\eta^2}{\delta^2}}\,\right)\\ &\times \left(\sgn \left[h_z(\bs{k})\right]\bigg|_{\bs{k} = D_1} - \sgn \left[h_z(\bs{k})\right]\bigg|_{\bs{k} = D_2} \right).
\end{split}
\label{eq:formula}
\ee 
This relation is valid for all the models discussed in this paper. Besides the constant term without any $\bs{k}$-dependency, the above expression clearly shows that the Chern number will be determined by the values of $h_z(\bs{k})$ at two Dirac points. This relation also predicts that the possible values of the Chern number are $C = 0,\,\pm1$. 

We have summarized the calculation of the Chern numbers using the analytical expression given in Eq. \eqref{eq:formula} in Table \ref{table} for all the models of 2D $N$-stacked SSH chains studied in Secs. \ref{sec:Case-I}, \ref{With-TR}, and \ref{Without-TR}. We see from the table that, for the nontrivial topological cases with $C\neq 0$, a square root factor appears from the {\it mass} term of Eq. \eqref{eq:formula}, which is identical to the first term of Eq. \eqref{eq:formula}. Consequently, the square root disappears in the expression of the Chern number, and the condition $|\delta| > |\eta|$ gives $\sgn[1-(\eta/\delta)^2] = +1$.

\subsection{Phase diagram}

We now concentrate only on those two models, which showed nontrivial topology with $|C| = 1$, to study their topological phase transitions. Here, we fix the parameter $\eta = 0.5$, which decides the relative strength of the intra-dimer and inter-dimer hopping within an SSH chain. We then investigate the system's phase transition by tuning the system parameters $(\delta,\,\gamma)$, where these parameters decide the hopping strength between two neighboring SSH chains. In Figs. \ref{Fig-Phase-diag} (a) and (b), we present the phase-diagrams on the parameter space $(\delta,\,\gamma)$ of the Model 2 and 3 of Sec. \ref{Without-TR}. Here we also relax any restriction on the values of $(\delta,\,\gamma)$: these parameters can be both positive and negative. Since we set $\eta = 0.5$, the Dirac points can only exist if $|\delta| \geq 0.5$. Consequently, when $|\delta| < 0.5$, the systems should show a trivial topology with $C=0$. In Fig. \ref{Fig-Phase-diag}, we indeed see for both the systems that the Chern number $C=0$ in the region $|\delta| < 0.5$ of the parameter space and this region is highlighted by light-grey color. The phase diagrams show that, for these two models, the transition from trivial to nontrivial occurs when $\delta \geq \eta$. These figures also reveal that the topological properties of the two models are complementary to each other, i.e., wherever in the parameters space, the Chern number of one system is $C$, and the Chern number of the other system is $-C$. The parameter regions with nonzero Chern numbers are shown using cyan and grey colors. 

\section{\label{sec:ZeroCN} Cases of the nontrivial topology with the Chern number $C=0$}

This section focuses on the $C=0$ cases observed in Secs. \ref{sec:Case-I},  \ref{With-TR}, \ref{Without-TR}. The basic model, which was considered at the beginning, is anisotropic. The anisotropy appears due to the independent intra-chain and inter-chain (both vertical and diagonal inter-chain bonds) hopping strength. This type of 2D anisotropic model can show nontrivial topological properties even though the Chern number of a system is {\it zero} \cite{ZeroChernNo}. For the anisotropic 2D model with $C = 0 $, the topological invariant is the 2D Zak phase, which is nothing but the 1D Zak phase with fixed values of momenta. The 2D Zak phase is defined as
\begin{equation}
Z_{x/y, m}(k_{y/x}) = \int_{BZ} A_{x/y, m}(k_{x}, k_{y}) \, dk_{x/y},
\label{Eq-2D-Zak}
\end{equation}
where $m$ represents the band indices for lower (-) and upper band (+). The Berry connection $A_{x/y, m}(k_x, k_y)$ is defined in \cite{Note}. 

This investigation begins with the model presented in Sec. \ref{sec:Case-I}. This model has all three previously mentioned fundamental symmetries. For this model, we find the 2D Zak phase $Z_{x,-}(k_y) = -\pi$, provided the $k_y$ values are within the range given in Eq. (\ref{Eq-k_y-cond}). However, for $k_x = \pi$, the 2D Zak phase $Z_{y,-}(k_x) = 0$. These values of the Zak phase are the same for both trivial and nontrivial individual SSH chains. 

\begin{figure}[t]
\includegraphics[width=8.5cm,height=3cm]{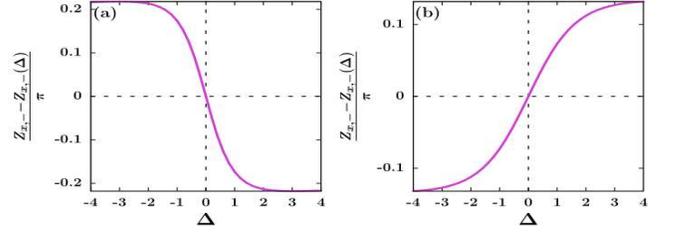}
\caption{The difference between the 2D Zak phases of the basic model (Fig. \ref{Fig-Primary_Model}) and the chiral symmetry broken system [Fig. \ref{Fig-E-Model_PT}(a)] is shown as a function of the NNN hopping amplitude $\Delta = 2 \gamma$. In subfigures (a) and (b), the individual SSH chain with trivial and nontrivial topology is considered, respectively. These figures demonstrate the presence of the fractional Zak phase for the chiral symmetry broken 2D $N$-stacked SSH model [Fig. \ref{Fig-E-Model_PT}(a)]. The Zak phases $Z_{x,-}$ and $Z_{x,-}(\Delta)$ are calculated at $k_y = \pi$.}
\label{Fig-frac-Zak-phase}
\end{figure}

\begin{figure}[b]
\includegraphics[width=8cm,height=5cm]{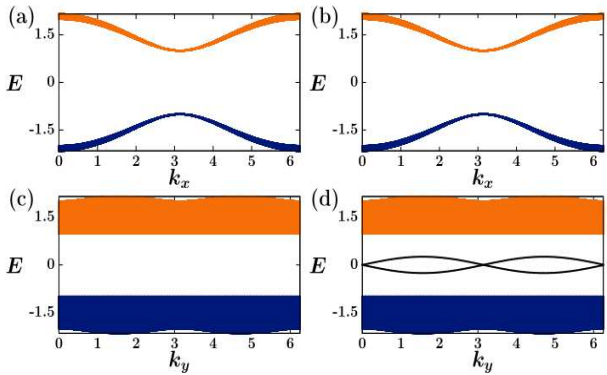}
\caption{The band diagram is presented for the model shown in Fig. \ref{Fig-Im-diag-y-hop}(a), where the strengths of the diagonal NNN and the vertical hopping are equal. In subfigures (a) and (b), we set $\eta = -0.5$ and $0.5$ so that the individual SSH chain will be trivial and nontrivial, respectively. Here, we choose $\delta = 0$ for both the cases.}
\label{Fig-C0}
\end{figure}

\begin{figure}
\includegraphics[width=8cm,height=5cm]{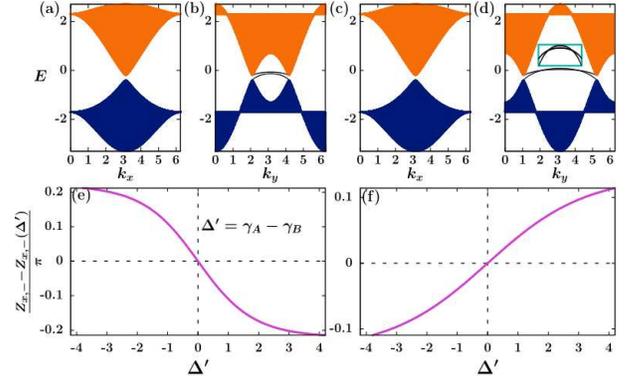}
\caption{The energy bands are shown under POBCs of the system presented in Sec. \ref{sec:Case-II}A, but now with the broken PH symmetry. Since here both chiral and PH symmetries are broken, the energy bands are asymmetric. The Chern number is $C=0$ in this case. Subfigures (a) and (b) present the results when the individual SSH chain is trivial ($\eta = -0.5$); whereas subfigures (c) and (d) are shown for the case when the individual SSH chain is non-trivial ($\eta=0.5$). Here we set the parameter $\delta=1.0$ for both the cases. Their corresponding deviation from the Zak phase of the basic model is presented respectively in subfigure (e) and (f). In subfigures (a)-(d), we set the hopping amplitude $\gamma_A = 0.2$ and $\gamma_B = 0.1$, which break the PH and inversion symmetries in the system.}
\label{Fig-PH-break-01}
\end{figure}

We then study the model presented in Sec. \ref{With-TR} that is an $N$-stacked SSH model with broken chiral symmetry. Because of the existence of the non-zero $h_z$, the pseudo-spinors move in the $z$-direction from the equatorial plane (represented by $h_z = 0$) of the Bloch sphere. This aberration of results is reflected in the non-quantized Zak phase, i.e., the Zak phase is not an integer multiple of $\pi$. This is referred to as the fractional Zak phase \cite{Zakphase-01, Zakphase-02, Zakphase-03, Zakphase-04}. The presence of the fractional Zak phase has also been experimentally observed for a chiral symmetry broken 1D TI \cite{Zakphase-01}. For our model, we demonstrate in Fig. \ref{Fig-frac-Zak-phase} the deviation of the Zak phase from the quantized to non-quantized values as a function of the NNN hopping amplitude $\Delta = 2\gamma$, the parameter which breaks the chiral symmetry. In Fig. \ref{Fig-frac-Zak-phase}(a), we show the deviation of the Zak phase for the case when the individual SSH chain is trivial, whereas Fig. \ref{Fig-frac-Zak-phase}(b) shows the same deviation when the individual SSH chain is nontrivial. The nature of the deviation of the Zak phase for these two cases is the opposite. According to our expectation, for both cases, when the parameter $\Delta\rightarrow 0$, the Zak phase approaches $-\pi$, a quantized value. Moreover, we also notice that the Zak phase saturates when $|\Delta| \geq 4$. However, in this parameter region, the NNN hopping strength becomes stronger than the NN hopping.

We now analyze the $C=0$ cases discussed in Sec. \ref{Without-TR}. Here, both chiral and TR symmetries are broken. This section considers two models with $C=0$: Model 1 and Model 4. The energy bands of Model 1 are similar to the basic model. Consequently, the Zak phase of Model 1 is equals the basic model. In the case of Model 4, when the inter-chain coupling parameters are equal (i.e., $\gamma_1 = \gamma_2 = \gamma$), the system exhibits $C=0$. According to Ref. (\cite{ZeroChernNo}), the nontrivial topology of this anisotropic system appears through the presence of gap-edge states, which are completely isolated from the bulk part. In Fig. (\ref{Fig-C0}), similar gap-edge states in the whole BZ are also observed in our model. Here again, we consider two cases: all the SSH chains are topologically nontrivial, and all are trivial. The respective energy band diagrams are shown in Figs. \ref{Fig-C0}(a)-(d). 
In these cases, we set $\delta = 0$. We show in Fig. \ref{Fig-C0}(d) that the system has gap-edge states. Therefore, we concentrate only on this case. We find the 2D Zak phase $Z_{x,-}(k_y) = -\pi$, provided the values of $k_y$ lie within the range given in Eq. (\ref{Eq-k_y-cond}). However, for $k_x = \pi$, the 2D Zak phase $Z_{y,-}(k_x) = 0$. Moreover, the area covered by the edge states in the BZ decreases after increasing $\delta$, which is shown in Fig. \ref{Fig-Im-diag-y-hop}(d)-(e).

\section{\label{sec:PH-broken}Systems with broken particle-hole symmetry}

\begin{figure*}[t]
\includegraphics[width=15cm,height=7cm]{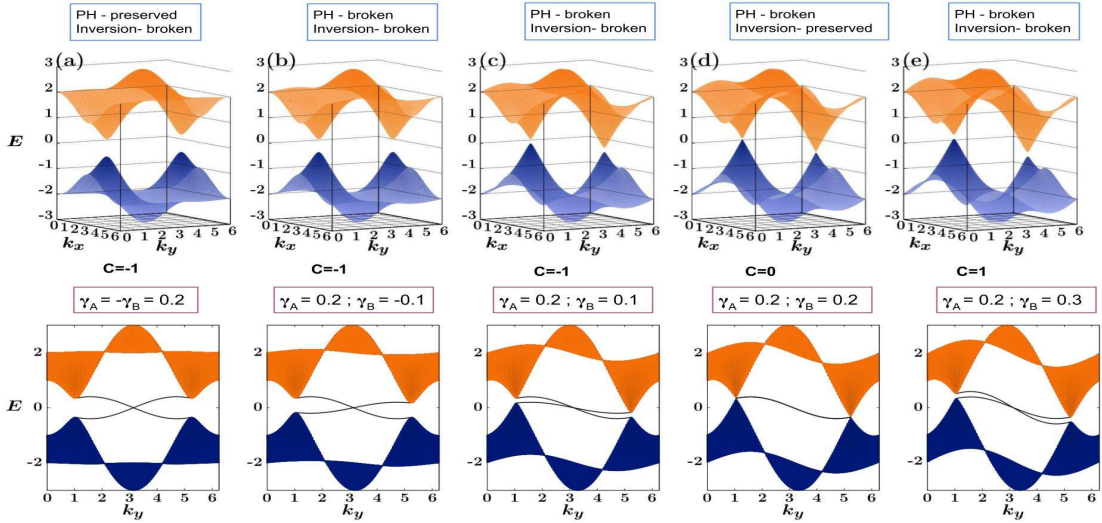}
\caption{The energy bands are shown under POBCs with the broken PH symmetry of {\it Model 3} discussed in Sec.\ref{sec:Case-II}B. In the upper panel, energy $E$ is plotted as a function of the momenta $k_x$ and $k_y$ for different values of the intra-sub-lattice hopping amplitudes $\gamma_A$ and $\gamma_B$ for {\it Model 3} (discussed in Sec. \ref{sec:Case-II}B). The corresponding band diagram under POBCs with periodic boundary condition along $y$-direction are shown in the lower panel. As we gradually change the amplitudes $\gamma_A$ and $\gamma_B$, the energy bands become asymmetric. Although there is a topological phase transition point at $\gamma_A = \gamma_B = 0.2$, where the system has inversion symmetry; but the PH symmetry is still broken at this point. Around this transition point, the system makes topological phase transition from a phase with Chern number $C = -1$ to a phase with $C = 1$. Here we set the dimerization constant $\eta = 0.5$ and hence the individual SSH chain is topologically nontrivial.}
\label{Fig-PH-break-NTri}
\end{figure*}

\begin{figure*}[t]
\includegraphics[width=15cm,height=7cm]{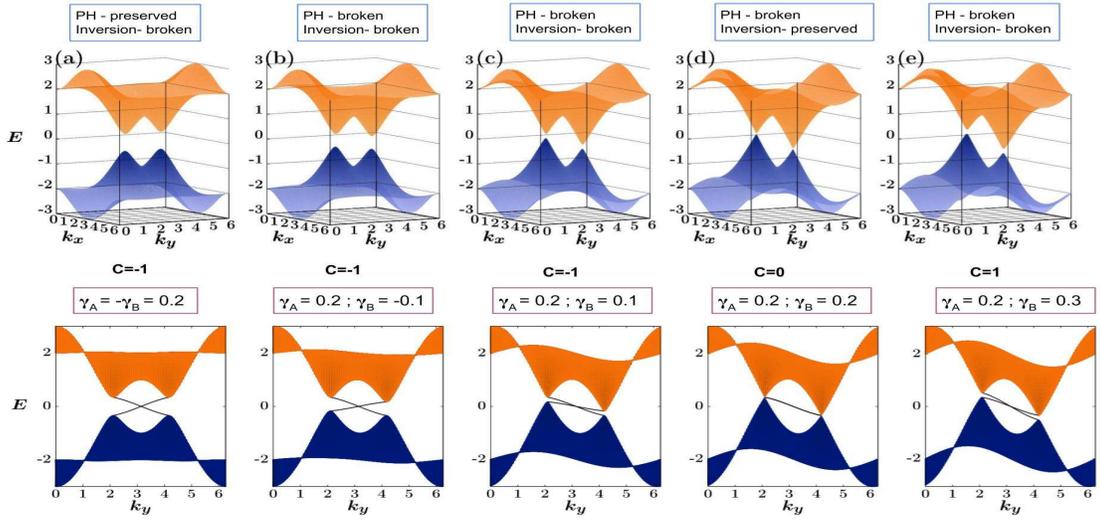}
\caption{The same results are presented as in Fig. \ref{Fig-PH-break-NTri}, but here we set the dimerization parameter $\eta = -0.5$ to make the individual SSH chain topologically trivial.}
\label{Fig-PH-break-Tri}
\end{figure*}

We now discuss the PH symmetry broken version of the models presented in Secs. \ref{sec:Case-II}A and \ref{sec:Case-II}B. In Fig. \ref{Fig-PH-break-01}, the energy bands are presented for the PH symmetry broken version of the model presented in Sec. \ref{sec:Case-II}A under POBCs. Here, subfigures \ref{Fig-PH-break-01}(a) and \ref{Fig-PH-break-01}(b) are plotted when the individual SSH chain is trivial; whereas subfigures \ref{Fig-PH-break-01}(c) and \ref{Fig-PH-break-01}(d) are presented for the cases, when the individual SSH chain is nontrivial. In order to break the PH symmetry in the model, we set the $A$ to $A$ hopping amplitude $\gamma_A$ and the $B$ to $B$ hopping amplitude $\gamma_B$ unequal. Here, we particularly consider $\gamma_A = 0.2$ and $\gamma_B=0.1$. Our results remain qualitatively invariant for any other pairs of unequal values of these above mentioned hopping parameters. The broken PH symmetry in the model leads to the asymmetric energy bands. In this case, we obtain the Chern number $C=0$ as we obtained for the model presented in Sec. \ref{sec:Case-II}A, where only chiral symmetry was broken. Again like the chiral symmetry broken case, here we obtain the fractionalized Zak phase for this case. 
For the equal hopping amplitudes $\gamma_A = \gamma_B = 0.2$, the Zak phase gets quantized due to the preserved inversion symmetry. In order to preserve the inversion symmetry, the Hamiltonian should satisfy the following condition: 
\be
\mathcal{R}^{-1} \, \mathcal{H}(k_x,k_y) \, \mathcal{R} = \mathcal{H}(-k_x,-k_y),
\ee
where $\mathcal{R}$ represents the inversion operation. The deviations of the Zak phase from the Zak phase of the basic model with all the symmetries preserved are shown in Fig. \ref{Fig-PH-break-01}(e) (individual SSH is trivial) and (f) (individual SSH is nontrivial). Here the deviation of the Zak phase is plotted as a function of the difference $\Delta^\prime$ between two intra-sub-lattice hopping amplitudes, i.e., $\Delta^\prime = \gamma_A - \gamma_B$.

We then investigate the effect of the PH symmetry breaking on the models presented in Sec.\ref{sec:Case-II}B. Here, we present only {\it Model 3} of Sec. \ref{sec:Case-II}B, because the other models of Sec. \ref{sec:Case-II}B show qualitatively similar behavior. The band diagram for the {\it Model 3} under full PBC and POBCs are shown in Figs. \ref{Fig-PH-break-NTri} and \ref{Fig-PH-break-Tri}. The subfigures \ref{Fig-PH-break-NTri}(a)-(e) show the band diagrams corresponding to the nontrivial SSH chain; and on the other hand, the subfigures \ref{Fig-PH-break-Tri}(a)-(e) show the band diagrams corresponding to the trivial SSH chain. In the upper panel, the band diagrams are plotted under full PBC and their corresponding band diagrams under POBCs are shown in the lower panel.  Here, we observe a topological phase transition from the topological phases with the Chern number $C = -1$ to $C = +1$ due to the interplay of the breaking and preserving of the PH and the inversion symmetries. The topological phase transition occurs in the system when the system has inversion symmetry with equal hopping amplitudes $\gamma_A = \gamma_B = 0.2$.

\section{\label{sec:Exp-aspects}Experimental aspects of 2D $N$-stacked SSH model }

We now discuss the possible experimental realizations of the 2D $N$-stacked SSH model. In order to construct the 2D $N$-stacked SSH model, we first need to construct single 1D SSH chain. Recent discoveries suggest that the SSH chain could be realized in photonics \cite{Exp-01} and electric circuits \cite{Exp-07}. In photonics, one-dimensional TIs can be engineered using an array of coupled optical waveguides \cite{Exp-01, Exp-02, Exp-04, li2014topological}. In Ref. \cite{Exp-01}, to design a 1D SSH chain, all the waveguides are kept at the same distance from each other and an auxiliary waveguide is placed next to every other waveguide. These auxiliary waveguides strengthen the coupling between the two waveguides that are same distance apart and thus simulate the intra-cell and inter-cell hopping in the 1D SSH chain. The array of optical waveguides can be fabricated using a femtosecond laser writing technique \cite{Exp-01, Exp-02, Exp-03}. Moreover to design the SSH chain in the electric circuits, each sub-lattice is constructed from a combination of inductors and capacitors.

Similarly using the optical waveguides, the E-SSH model (mentioned in the Introduction section) with broken chiral symmetry can also be realized \cite{Exp-04}. This fabrication uses photon propagation along a binary waveguide lattice placed in a zig-zag geometry \cite{Exp-04}. In this experiment, the hopping is considered in such a manner that the system breaks chiral symmetry, but possesses inversion symmetry. The inversion symmetry in the system can be preserved by keeping the equal hopping amplitudes between A to A sub-lattice and B to B sub-lattice. This design is helpful in breaking the symmetries of the SSH chain. It is also studied that 2D photonic crystals with broken TR symmetry can be realized using photonic meta-materials \cite{Haldane2008}.

Recent studies also have found that a 2D $N$-stacked SSH chain can also be fabricated using magneto-mechanical meta-material \cite{Exp-05} and electric circuits \cite{Yang2022, Exp-07}. In \cite{Exp-05}, the 2D $N$-stacked SSH model is realized with dislocation defects, where the 1D SSH chain is designed using mechanical resonators. These mechanical resonators are coupled to each other through their magnetic interaction \cite{Exp-05, Exp-06}. Each resonator is like a sub-lattice and a unit cell is made of two coupled adjacent resonators. A 2D $N$-stacked SSH model is designed when this setup is aligned in the $x$-direction and kept periodically along the $y$-direction.

Here we propose that a 2D $N$-stacked SSH model can be fabricated by designing 1D SSH chain using the techniques mentioned above and place them periodically in the transverse direction of the SSH chains. The chiral symmetry can be broken following the method given in \cite{Exp-04}. Moreover, imaginary hopping can be introduced (to break the TR symmetry) by applying a magnetic field-like gauge field in the direction perpendicular to the plane of the 2D lattice. Since the chiral, PH, and inversion symmetries are dependent on the hopping amplitudes from $A$ to $A$ and from $B$ to $B$ sub-lattices, these symmetries can be broken by properly placed the waveguides.

\section{\label{sec:Final Remarks} Summary}

In this paper, we have proposed a 2D SSH model constructed by stacking $N$ number of SSH chains with coupling only between two neighboring SSH chains. Here, we have considered two versions of the model determined by the topological property of the individual SSH chain: (1) all the chains are topologically trivial with the winding number $w=0$; and (2) all the chains are topologically nontrivial with $w\neq 0$. Depending on the intra-chain parameters (intra-dimer and inter-dimer strengths) and the inter-chain coupling strength, the 2D $N$-stacked SSH chains can be in three phases: topologically trivial, topological semimetal, and weak topological insulator. We have observed all these phases in this model by systematically breaking the symmetries.  

We start the analysis with a basic model of the $N$-stacked SSH chains with all the fundamental symmetries (chiral, TR, and PH) preserved. Depending on the topological property of the individual chain, the basic model shows both topologically trivial and nontrivial nature. However, for both the cases, the Chern number of this model $C=0$. The nontrivial topology of this model is identified from the presence of the edge states. We then start analyzing the role of different symmetries on the topology of this model.

First, we break the chiral symmetry by introducing intra-sub-lattice hopping in the system. This is done by allowing NNN hopping within an SSH chain. The introduction of this new hopping leads to a mass-like term, and consequently, it opens up the band gap. This model also shows the Chern number $C=0$. However, edge states are still present in this model, which indicates its nontrivial topology. 

Following Haldane, we have broken the TR-symmetry by introducing  intra-sub-lattice hopping strength with imaginary amplitude. For this case, we have studied four different models. Only two models (Model 2 and Model 3) have shown nontrivial topology with $C = \pm 1$. The other models, Model 1 and Model 4 have the Chern number $C=0$, but the presence of edge states in theses models indicate their nontrivial topology. However, Model 4 shows the nontrivial topology with a nonzero Chern number when two different types of inter-chain hopping strengths are considered unequal. In all these cases, we have preserved the PH symmetry in all the models. Therefore, we could apply a recently proposed analytical formulation to calculate the Chern number. This calculation agrees well with the models' numerically observed topological properties. We have presented phase diagrams of these systems that showed nontrivial topology with nonzero Chern numbers. Thus we have found a recipe to prepare a Chern insulator from a weak TI with $C = \pm1$ and cataloged the topological phases of the 2D $N$-stacked SSH chains by the systematic breaking of symmetries.

Next, we have concentrated separately on studying the nontrivial topological cases with $C=0$. We have calculated the 2D Zak phase as a topological invariant for these cases. Interestingly, here we have observed two different behaviors in the 2D Zak phase: quantized Zak phase, which is equal to the integer multiples of $\pi$, and fractional Zak phase when this is not an integer multiple of $\pi$. The former is observed in the basic model (discussed in Sec. \ref{sec:Case-I}), and also in chiral and TR symmetry broken Model 1 and Model 4 (discussed in Sec. \ref{sec:Case-II}B). The latter is observed when the chiral symmetry is broken in the basic model (discussed in Sec. \ref{sec:Case-II}A). For this model, we have extensively discussed how the breaking of the chiral symmetry affects the fractional nature of the Zak phase. 

Finally, we have presented the results for the system with broken PH symmetry for the two cases: with broken chiral symmetry, and with broken chiral as well as the TR symmetry. The breaking of the PH symmetry introduces an asymmetry in the hopping terms of the system, which exhibits in the energy bands and also in the phase transition around the point where the system preserves inversion symmetry. We have also discussed possible experimental realizations of the 2D $N$-stacked SSH model.

\section{\label{sec:Outlook} Outlook}

Three-dimensional weak topological insulators (WTIs) formed by stacking 2D QSH layers have been studied extensively, both theoretically and experimentally. These studies also consider the impact of breaking of some of the fundamentals and translational symmetries \cite{Pauly2015,oh2023ideal,Huang_2023,Ran2009,Zhang2021,Fan2017,PhysRevB.107.024506,PhysRevB.104.104510,Noguchi2019,PhysRevB.107.024506,Anirban2023, LIU2012906, Du2021,NQHE,PhysRevB.107.L041402,PhysRevB.85.165110,Kim2020,WTI-symm}. However, compare to 3D WTIs, there are relatively a few theoretical and experimental studies available on the 2D WTIs constructed from the stacked 1D TIs \cite{WeylSSH-01,WeylSSH-02,WeylSSH-03,Exp-05,Yang2022, Exp-07}. In this extensive paper, we have introduced five different models (presented in Secs. \ref{sec:Case-II}A and \ref{sec:Case-II}B) which can be studied further with the broken translational symmetry also.

As a future perspective, one can construct a full-fledged three-dimensional layered structure by extending the 2D $N$-stacked SSH model. In its 3D version of the 2D $N$-stacked SSH model, one can place all the 2D $N$-stacked layers periodically along the $z$-direction. This extension of 1D TIs to 3D WTIs will also exhibit following exotic topological properties as observed in the 3D layered QSH system: such as non-linear QHE, spin-polarization, Quantum anomalous layer Hall effect, QSH effect with half integer, etc. \cite{Anirban2023,LIU2012906,NQHE,Du2021}. This work is useful in choosing a model, depending on the requirement of the study, which either preserve or break the fundamental symmetries. For this 3D extension, one can also look into the macroscopic electric, magnetic and optical properties via symmetry breaking.

\begin{acknowledgments}
Authors acknowledge financial support from DST-SERB, India, through the Core Research Grant CRG/2020/001701. One of the authors (JNB) acknowledges Max Planck Institute for the Physics Complex Systems (MPI-PKS), Dresden, Germany, for their hospitality and giving an opportunity to discuss some of the experts in the field. Particularly, his interaction with Dr. Mohsen Yaarmohammadi, currently at UT-Dallas, was beneficial. Authors also like to thank the anonymous referees for their valuable comments, which helped in improving the manuscript.
\end{acknowledgments}

%\bibliography{References}

%

\end{document}